# High-Linearity PAM-4 Silicon Micro-ring Transmitter Architecture with Electronic-Photonic Hybrid DAC

Zheng Li[1,2,*], Chengyang Lv[1,3,*], and Min Tan[1,3†]

[1] *School of Integrated Circuits, Huazhong University of Science and Technology*
[2] *School of Optical and Electronic Information, Huazhong University of Science and Technology*
[3] *Wuhan National Laboratory for Optoelectronics, Huazhong University of Science and Technology*
[*]These authors contributed equally
[†] Corresponding author: mtan@hust.edu.cn

## Abstract

This paper presents a high linearity PAM-4 transmitter (TX) architecture, consisting of a three-segment micro-ring modulator (MRM) and a matched CMOS driver. This architecture can drive a high-linearity 4-level pulse amplitude (PAM-4) modulation signal, thereby extending the tunable operating wavelength range for achieving linear PAM-4 output. We use the three-segment MRM to increase design flexibility so that the linearity of PAM-4 output can be optimized with another degree of freedom. Each phase shift region is directly driven by the independently amplitude-tunable Non-Return-to-Zero (NRZ) signal. The three-segment modulator can achieve an adjustable wavelength range of approximately 0.037 nm within the high linearity PAM-4 output limit when the driving voltage varies from 1.5 V to 3 V, simultaneously achieving an adjustable insertion loss (IL) range of approximately 2 dB, roughly four times that of the two-segment MRM with a similar design. The driver circuit with adjustable driving voltage is co-designed to adjust the eye height to improve PAM-4 linearity. In this article, the high linearity PAM-4 silicon micro-ring architecture can be employed in optical transmitters to adjust PAM-4 eye-opening size and maximize the PAM-4 output linearity, thus offering the potential for high-performance and low-power overhead transmitters.

## I. INTRODUCTION

As artificial intelligence and machine learning continue to advance rapidly, data centers are faced with an unprecedented demand for increased data throughput [1]. Silicon Photonics (SiPho) has emerged as a key enabling technology for optical interconnects in the next generation of data centers [2], [3]. The Mach-Zehnder modulator (MZM) holds a significant share of the market for external optical transmitter modulators due to its high bandwidth and process reliability. However, due to the exceedingly rapid growth in data communication demands, the energy consumption of transceiver links accounts for a substantial portion of data center operations [4]. In contrast, the MRMs exhibit low energy consumption in driving and have a small footprint, making them suitable for potential large-scale integration, which indicates greater potential compared to the MZM [5]. Recent demonstrations have indicated that silicon MRMs can achieve data rates of 240 Gb/s using PAM-4 [6] and 330 Gb/s using PAM-8 [7]. This underscores the capability of MRM to meet the requirements of the upcoming generation of 1.6 Tb/s optical modules. Nevertheless, MRM is susceptible to process variations, noise, and temperature effects, leading to performance degradation [8]. In the presence of the Lorentzian response of the MRM, achieving a linear output for equidistant PAM-4 signal levels is also a challenging problem.

Due to the nonlinearities inherent in MRM-based intensity modulation, the practicality of employing multi-level modulation schemes such as PAM-4 is constrained. Currently, two primary methods are employed for generating high linearity PAM-4 optical signals, utilizing an electronic digital-to-analog converter (DAC) to drive for generating signals and optical DAC with multi-segment phase shift region, respectively. Electronic DAC implementation involves a relatively intricate design of an electronic integrated circuit (EIC), incorporating components such as a PAM-4 pattern generator, serializer, and MRM driver. Additionally, to compensate for the impact of static and dynamic MRM nonlinearity, it may be necessary to design a pre-distortion circuit to achieve high linearity PAM-4 output [9], [10]. This approach may heighten the design complexity and energy consumption of the EIC. Optical DAC implementation involves a two-segment MRM comprising a longer phase shifter for the most significant bit (MSB) and a shorter phase shifter for the least significant bit (LSB). The generator's odd and even bits are serialized separately and serve as inputs to independent MSB/LSB output drivers [11], [12], [13]. Due to the requirement of controlling three distinct eye heights for PAM-4 signal output, while MRM has only two phase-shifting segments, it is susceptible to nonlinear effects, resulting in reduced linearity, degraded eye diagrams, and increased bit error rates. The three-segment MRM with three phase shifters of different lengths has previously been proposed. However, its driving method wastes the length of phase shifters, and the modulation efficiency is relatively poor [14], [15]. In addition, achieving high linearity PAM-4 output can also be realized through multi-segment MRMs [16][16] or cascaded MRMs [17]. However, these approaches undoubtedly increase the complexity of device design and fabrication.

In this paper, we introduce a high linearity PAM-4 silicon micro-ring transmitter architecture with electronic-photonic hybrid DAC, consisting of a three-segment MRM with three phase shifters of the same length and driver circuits with independently adjustable output voltage through co-design of optical and electronic circuit, offering enhanced flexibility in MRM design for linear PAM-4 output for optical transmitters. This structure breaks the trade-off design between optical DAC and electronic DAC in terms of PAM-4 output linearity and driver circuit complexity. Each phase shift region is directly driven by NRZ signal generation circuits, which not only avoids increasing circuit design complexity and introducing additional power consumption but also enhances design flexibility. Through theoretical analysis and simulation, we verify this architecture widens the working wavelength range for PAM-4 linear output and increases the adjustable range of IL.

The rest of this paper is organized as follows: Section II provides a review of various PAM-4 output schemes, comparing them to discern the advantages of the transmitter architecture with the electronic-photonic hybrid DAC we designed. In Section III, we delve into the detailed design aspects of both the photonic integrated circuit and the electronic integrated circuit. Finally, our conclusions are presented in Section IV.

## II. COMPARISON OF PAM-4 OUTPUT SCHEMES

### A. Single-segment MRM (Electronic DAC)

To achieve PAM-4 optical signal output through the electronic DAC method, MRM with the single-segment shifter and the corresponding driver circuit is typically required. To achieve symmetric PAM-4 eyes, addressing the nonlinearity of MRM through nonlinear analog circuits or DSP+DAC technology is highly power-hungry. Similarly, employing predistortion methods also necessitates many stacked inverter linear NRZ driver segments, introducing additional power consumption [9], [10]. Given the significant power consumption in today's data centers with exploding data throughput, such high power consumption is not the best solution for the low-cost transmitter.

### B. Two-segment MRM (Optical DAC)

In contrast to the single-segment MRM driven by PAM-4 electronic signals, the two-segment MRM features two phase shift regions, each driven by independent electric NRZ radio frequency (RF) signals, to achieve optical PAM-4 signal output [11]. In the case of the one-segment MRM, although it simplifies the MRM structure design, relatively complex circuit designs will incur additional power overhead. On the other hand, the two-segment MRM is driven by two independent NRZ voltages to produce PAM-4 output, offering greater flexibility in designing energy-efficient driving circuits and achieving linearity. Regarding the MRM transmission curve, which exhibits a Lorentzian shape due to its nonlinearity, recent demonstrations have revealed that the two-segment MRM, with an optimized 1.9:1 MSB: LSB ratio, yields the best PAM-4 output [11].

Fig.1 shows the structure of the two-segment MRM and its driving scheme. However, due to the Lorentzian-shaped transmission curve, achieving linear PAM-4 output is confined to a narrow operating wavelength range. This limitation presents challenges in operating wavelength selection, resulting in a nearly fixed and non-adjustable IL. With the rapid growth in data communication volume, it becomes imperative to reduce the IL of transceiver devices while ensuring the quality of PAM-4 eye diagrams. Previous experiments have demonstrated that the power consumption of lasers constitutes over half of the total link power consumption [18]. In essence, under the same receiver sensitivity conditions, reducing the IL of the MRM can decrease the requirement for laser input power, which is a key factor in reducing overall link power consumption.

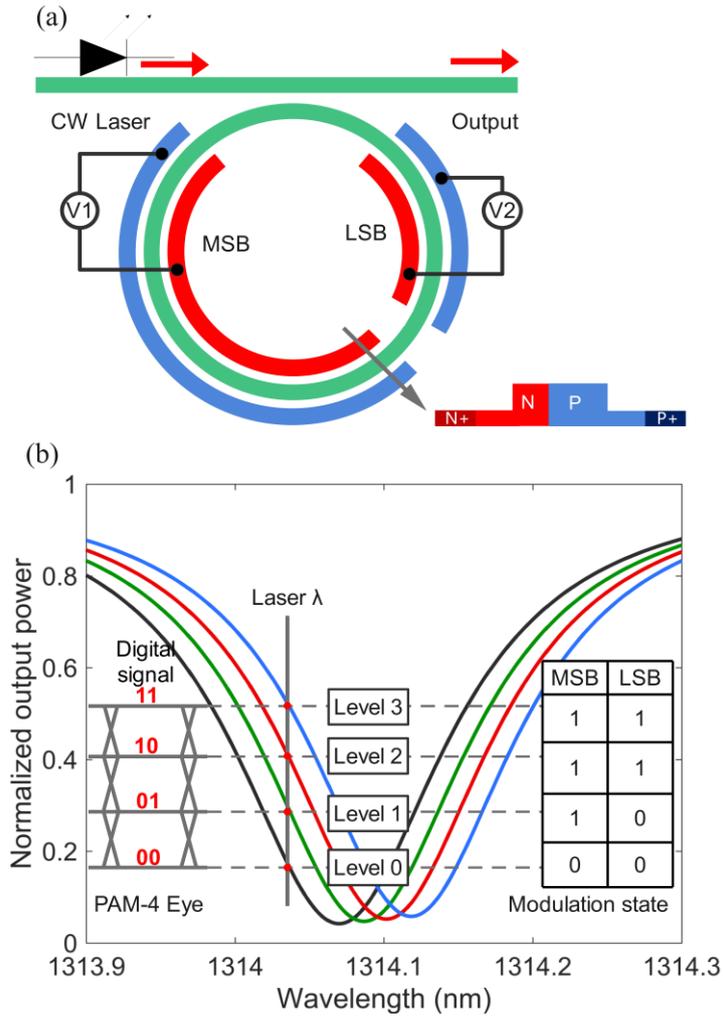

Fig. 1.  Overview of the two-segment MRM: (a) Top and cross-section views of the two-segment MRM. (b) Transmission of the two-segment MRM and its driving scheme when realizing PAM-4 output.

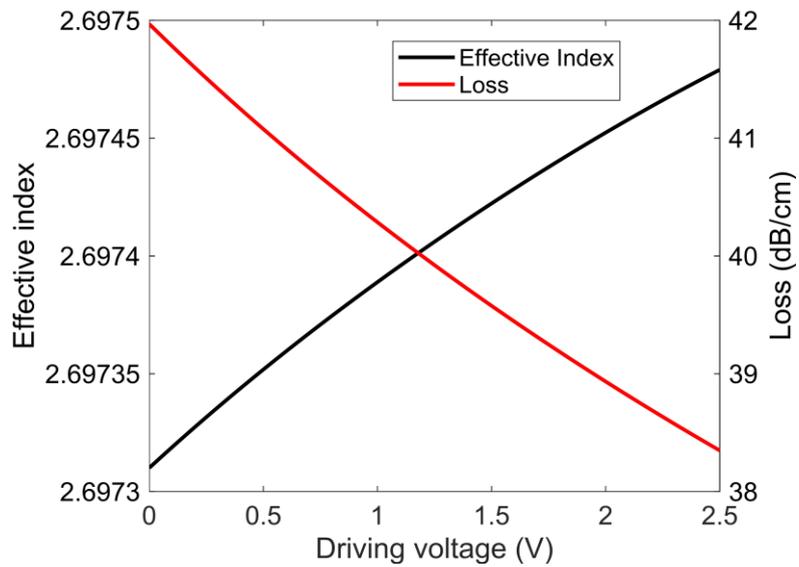

Fig. 2. The variation of effective refractive index and loss of active waveguide with respect to driving voltage.

Unfortunately, the PAM-4 optical output is constrained by the nonlinearity of MRR. According to Fig.3, the PAM-4 output linearity can be measured as PAM-4 ratio of level mismatch (RLM) [19]. The RLM is given by

$$P_{min}=(P_0+P_3)/2 \quad (1)$$
$$ES_1=(P_1-P_{min})/(P_0-P_{min}) \quad (2)$$
$$ES_2=(P_2-P_{min})/(P_3-P_{min}) \quad (3)$$
$$RLM=\min\{(3\times ES_1), (3\times ES_2), (2-3\times ES_1), (2-3\times ES_2)\} \quad (4)$$

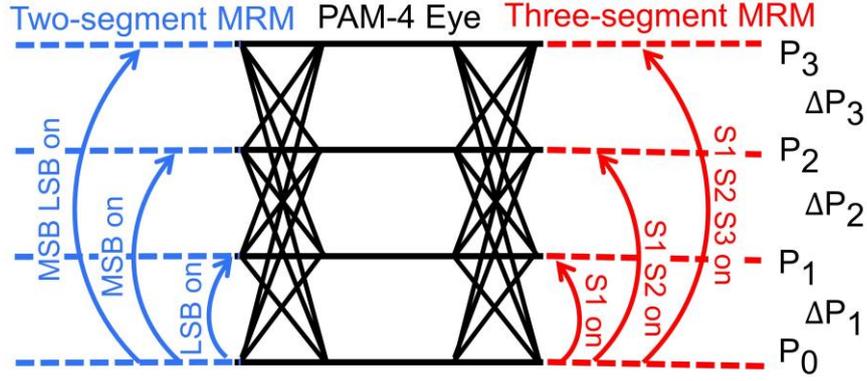

Fig. 3. PAM-4 eye diagram of two-segment MRM and three-segment MRM, which shows eye opening heights and their controlled variables.

Achieving an RLM value closer to 1 indicates better eye linearity. However, the two-segment MRM has only two independently controlled phase shifters but needs to control three eye heights, which introduces challenges in controlling the linearity of the PAM-4 eye. As is shown in Fig.3, eye-opening height $\Delta P_1$ is controlled by the voltage applied to LSB, whereas eye-opening height $\Delta P_2$ is controlled by the voltage applied to MSB. If achieving linearity in the lower two eyes ($\Delta P_1 = \Delta P_2$) by adjusting the voltages $V_1$ and $V_2$, the eye-opening height $\Delta P_3$ becomes fixed rather than adjustable because no more variables are available for adjustment.

Despite the design of the two-segment MRM with a length ratio of 1.9:1 for its two phase shifters, achieving complete PAM-4 output linearity remains challenging due to the presence of only two independently controlled driving voltages. According to the PAM-4 output standard protocol, RLM > 0.95 is considered acceptable for the PAM-4 transmitters [19]. Even if the linear output of the PAM-4 signal is achieved at a specific operating wavelength, it is limited to an extremely narrow operating wavelength range due to the Lorentz-shaped transmission rather than the linear-shaped curve, resulting in a fixed IL for the device. In other words, the requirement for achieving PAM-4 linear output to reduce transmitter bit error rate (BER) imposes restrictions on the operational state of the two-segment MRM, including its IL and extinction ratio (ER). This inherent property presents challenges when attempting to adjust IL to lower overall link power consumption.

We verify the properties described above through the following simulation. The designed MRM is composed of a bus waveguide and a micro-ring resonator. Part of micro-ring resonator is PN junction designed to operate in depletion mode as the phase shifter. The vertical PN junction is designed that the P-type and N-type are asymmetrically distributed to optimize the modulation efficiency by maximizing the optical mode overlap with the depletion region of the PN-junction since the changing of hole concentration contributes a larger refractive index change [20]. The carrier distribution of the PN junction is simulated using the Lumerical device, and related parameters such as effective refractive index, group index, and loss were extracted using the Lumerical mode. Both the two-segment MRM and the three-segment MRM share a common phase shifter scheme, albeit with variations in the length and distribution of the phase shift zone. The variation of effective refractive index and loss of active waveguide with respect to driving voltage is depicted in Fig 2.

Fig.4 shows the properties of the two segments at different wavelengths, including transmission, IL, ER, RLM, and transmission penalty (TP). As a comprehensive measure of ER and IL indicators, TP is given by [21]

$$TP = 10\lg 2P_{IN}/(P_1 + P_0) \quad (5)$$

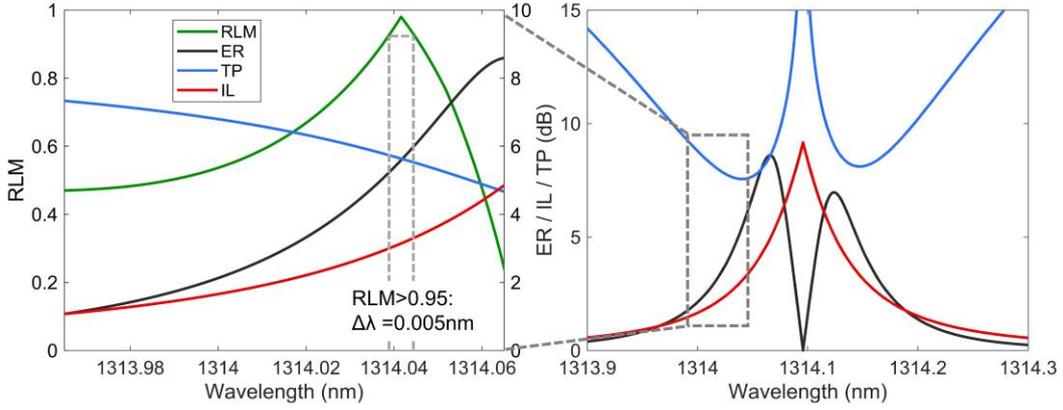

Fig. 4.  Properties of two-segment MRM when realizing PAM-4 output: The properties of IL, ER, TP of the two-segment MRM in O-band wavelength. The left part shows zoom-in of the properties and demonstrates PAM-4 output RLM.

where $P_1$ and $P_0$ are the power transmitted during a 1-bit and during a 0-bit respectively, and $P_{IN}$ is the input optical power.

The selection of the optimal operating point is, however, a trade-off between modulation bandwidth and modulation depth [22]. Our goal in choosing the operating wavelength is to achieve linear PAM-4 output reducing the BER of the transmitter. As Fig.4 shows, when the optical PAM-4 output is linear (that is, RLM>0.95), the corresponding operating wavelength range is merely 0.005nm (from 1314.039 nm to 1314.044 nm). The TP within this linear output wavelength range is from 5.57 dB to 5.69 dB; IL varies from 3.06 dB to 3.23 dB, and ER falls within the range of 5.39 dB to 5.83 dB. This illustrates that when ensuring the PAM-4 linear output of the transmitter, the operating wavelength should remain fixed within a narrow range, limiting the adjustment range of the two-segment MRM's IL to less than 0.2 dB. Although we have only discussed the case of 2V drive voltage, the result of changing the voltage is still not good, which will be discussed in more detail in the following section. Consequently, if we aim to select another operating wavelength to reduce the IL of the MRM while maintaining the same receiver sensitivity, changing the IL of the device becomes challenging due to the constraints imposed by PAM-4 output linearity.

*C. Three-segment MRM (Electronic-photonic Hybrid DAC)*

By contrast, the three-segment MRM has more freedom to adjust the operating wavelength in order to change the IL of the device, since it has three independent driving voltages added to three different phase shift zones respectively. Fig.3 shows the three controlled variables of three-segment MRM including S1, S2, and S3. By adjusting the driving voltages, we can change the operating wavelength under the condition that the optical PAM-4 output is linear so that we can change the IL and ER of the MRM in a relatively larger amplitude. Fig.5 (a) shows the structure of the three-segment MRM with three discrete phase shifter regions. Considering that the linearity of the PAM-4 output can be adjusted by the adjustable drive voltage of the matched CMOS driver designed by us at the same time and the design difficulty of MRM is reduced, so having three phase shifter regions of equal length is acceptable. Moreover, enough length of isolated area between two phase shifters should be guaranteed, since electric crosstalk in different phase-shift regions should be minimized. The length of three phase shift regions is designed to respectively account for 20% of the ring perimeter. Fig.5 (b) shows the transmissions of the three-segment MRM in different driving states and its driving scheme when realizing PAM-4 optical output. Just like the two-segment MRM, three-segment MRM has a similar phase shift region, which determines that the wavelength range is small when optical PAM-4 output is linear. For reasons of space, the analysis will not be repeated here. Differently, because of another degree of freedom, three-segment MRM has better properties to adjust the IL and ER when changing the driving voltages. We will analyze it in the next chapter.

### III. DESIGN OF HIGH LINEARITY PAM-4 SILICON MICRO-RING TRANSMITTER ARCHITECTURE WITH ELECTRONIC-PHOTONIC HYBRID DAC

In this chapter, we will extensively discuss the design of high linearity PAM-4 silicon micro-ring transmitter architecture with electronic-photonic hybrid DAC. Specifically, we will delve into the optical chip design of the three-segment MRM and co-designed voltage-tunable driver circuit.

*A. Photonic Integrated Circuit Design*

Fig. 5 shows the basic structure of the three-segment MRM. The MRM has a designed radius of 10 μm. We designed the proposed three-segment MRM based on the 220 nm SOI platform. The waveguide has a width of 420 nm with a slab thickness of 70 nm. 60% of the ring's circumference is designated as the phase shifter region, employing a depletion-type PN junction, consisting of three segments of identical length. The remaining part of the ring is undoped, serving as an isolated region to minimize electric crosstalk during modulation. Due to fabrication errors, achieving critical coupling for optimal ER in the three-segment MRM is challenging. Hence, in our simulations, we slightly over-couple the MRM to mimic a more realistic scenario.

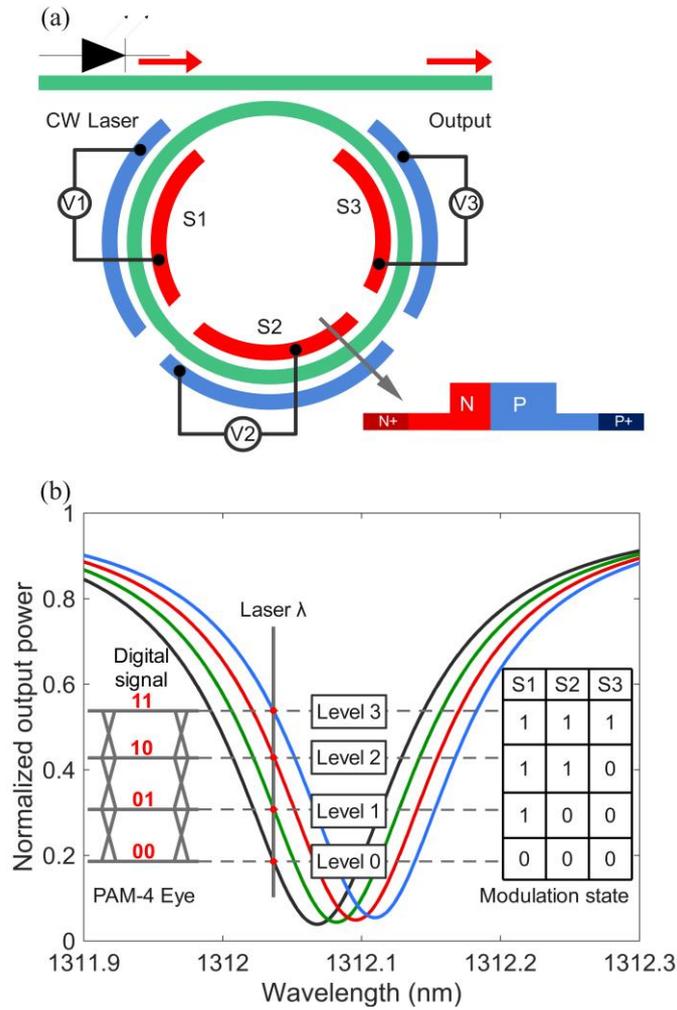

Fig. 5. Overview of the three-segment MRM: (a) Top and cross-section views of the two-segment MRM. (b) Transmission of the three-segment MRM and its driving scheme when realizing PAM-4 output.

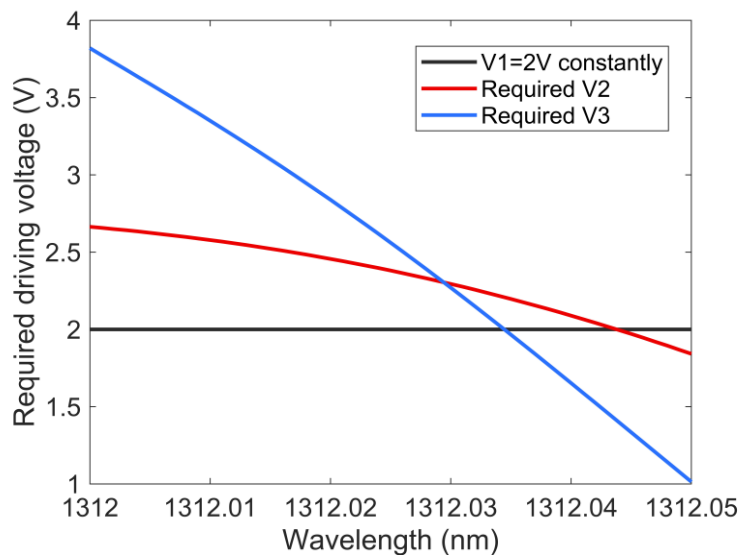

Fig. 6. The required voltages of $V_2$ and $V_3$ to realize linear PAM-4 output when $V_1$=2V (RLM=1).

The three-segment MRM offers greater flexibility in adjusting linearity across various operating wavelengths. To ensure a linear PAM-4 optical output (RLM = 1), we conducted tests on the required driving voltages $V_2$ and $V_3$ for $V_1$= 2 V. Fig. 6 shows the required voltages at different wavelengths across a 0.05nm range of different wavelengths (from 1314.1 nm to 1314.15 nm). As is shown in Fig.6, when a 2V voltage is applied to S1, the required $V_2$ and $V_3$ vary across diverse operating wavelengths. This helps us identify suitable voltages applied to distinct phase-shift regions, S2 and S3, thereby guiding the design of our electric driving circuit.

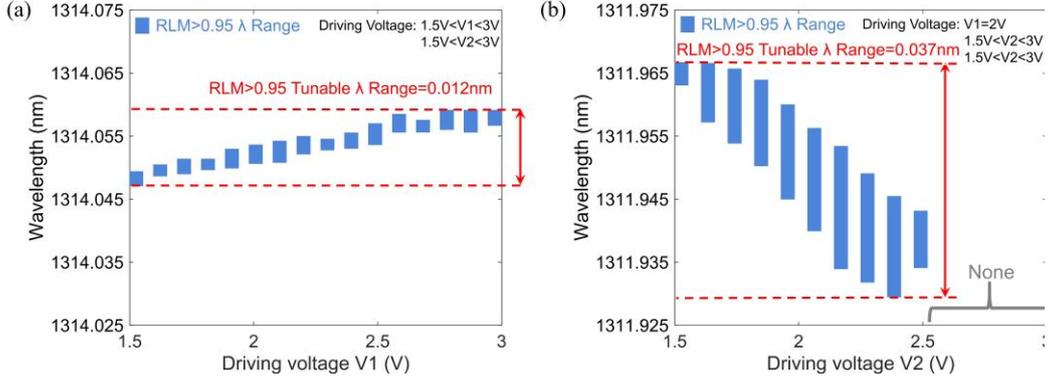

Fig. 7. The adjustable wavelength range of (a) two-segment MRM and (b) three-segment MRM when the PAM-4 output is linear (RLM>0.95) under the condition that the driving voltage is limited from 1.5V to 3V.

Considering the ease of designing a complementary metal-oxide semiconductor (CMOS) RF driving circuit to directly achieve low required drive voltages for co-packaging with the MRM chip, we assume the driving voltage is limited to the range from 1.5 V to 3 V. To demonstrate the advantages of the novel three-segment MRM, we analyze its adjustable operating wavelength range when achieving acceptable linear PAM-4 output (RLM>0.95). Furthermore, we compared this property between the two-segment MRM and the three-segment MRM. The wavelength ranges yielding linear PAM-4 output (RLM>0.95) are depicted in Fig.7 (a) and Fig.7 (b). As for the two-segment MRM, when the driving voltages of two segments are 1.5 V to 3 V arbitrarily adjustable, the wavelength range of linear PAM-4 output is limited to 0.012 nm (from 1314.047 nm to 1314.059 nm). We can see from Fig.1 (b) that the variation of corresponding IL is just about 0.5 dB (from 3.5 dB to 4.2 dB), that is, the linear PAM-4 output can be achieved by adjusting the driving voltage in this range. In contrast, the IL of the three-segment MRM exhibits greater flexibility in adjustment. Fig 7 (b) illustrates that with a preset driving voltage $V_1$ fixed at 2 V, $V_2$ and $V_3$ arbitrarily adjustable between 1.5 V to 3 V, like the previously mentioned driving voltage for the two-segment MRM, the adjustable wavelength range for linear PAM-4 output extends to 0.037 nm (from 1311.929 nm to 1311.966 nm), significantly larger than that of the two-segment MRM. This results in a considerably wider adjustable IL range of approximately 2.3 dB (from 2.6 dB to 3.9 dB) when realizing high linearity PAM-4 output, which can be observed in Fig.5 (b). This aspect is crucial for reducing the overall link loss of transmitters and subsequently decreasing link power consumption.

Overall, in comparison to the two-segment MRM, the three-segment MRM achieves a broader operating wavelength range while maintaining the requirement for linear PAM-4 output. This indicates a more flexible regulation of the IL without the expense of the loss of linearity of PAM-4 output.

## B. Electronic Integrated Circuit Design

Fig. 8 (a) shows the block diagram of the proposed 64 Gb/s PAM-4 three-segment MRM transmitter, consisting of a clock unit, a pseudo-random binary sequence (PRBS) generator, a retimer, a serializer, and three driver segments.

The architecture for quadrature clock generation is presented in Fig. 8 (b). The 8-GHz differential clock is first corrected by an AC-coupled inverter buffer with a feedback resistor, which converts the CML voltage level to full-swing CMOS logic [23]. An open-loop injection-locked quadrature (I/Q) clock generator (QCG) takes the corrected clock and generator four-phase quadrature clocks [24]. To overcome the phase and duty-cycle mismatches of the quadrature clocks due to process variation and unbalanced loading and routing, IQ correction (QEC) and duty-cycle correction (DCC) units are adopted [25].

The built-in pseudo-random bit sequence (PRBS) generator creates 8-wide parallel data streams which can be split into MSB and LSB branches for the data path. They are first retimed, after which the time-aligned data streams are encoded and serialized into 32-Gb/s NRZ data by the following Serializer to serve as inputs to independent three output drivers.

The block diagram for the PAM-4 Serializer is illustrated in Fig. 8 (c), comprising three slices. The 4-bit lookup table (LUT) array is configurable for NRZ/PAM-4 mode switching. In NRZ mode, all three LUT configurations employ identical encoding. In PAM-4 mode, thermal encoding is utilized in all three LUT configurations, as depicted in Fig. 9. The single-to-differential converter (S2D) array minimizes signal noise. The 4-1 MUX, based on tri-state inverters, employs a quarter-rate architecture. This design minimizes the clock tree requirements and enhances bandwidth [26], as depicted in Fig. 10.

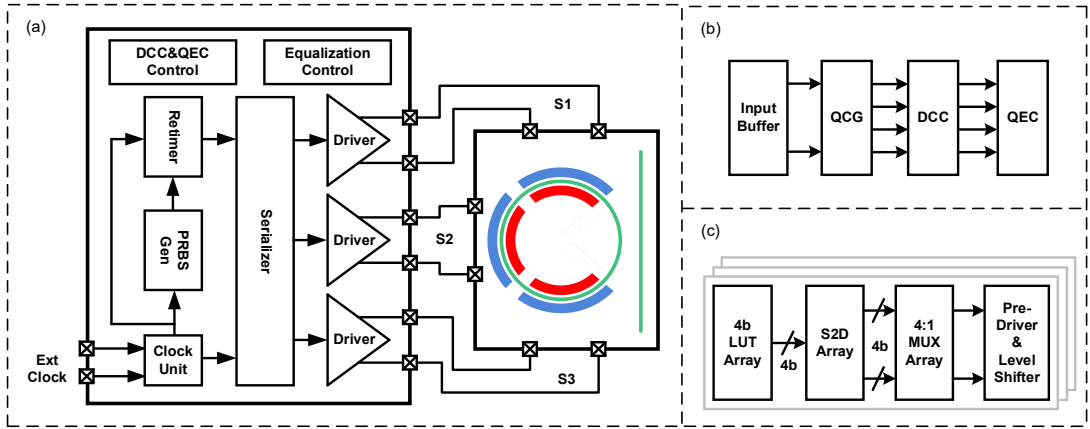

Fig. 8.  (a) Transmitter block diagram. (b) Quadrature clock generation architecture. (c) Serializer block diagram.

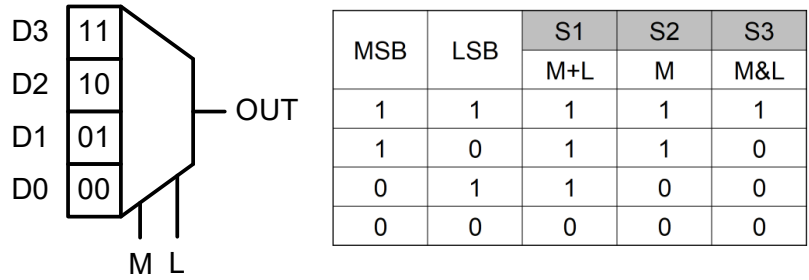

Fig. 9. LUT symbol and output mode configuration table.

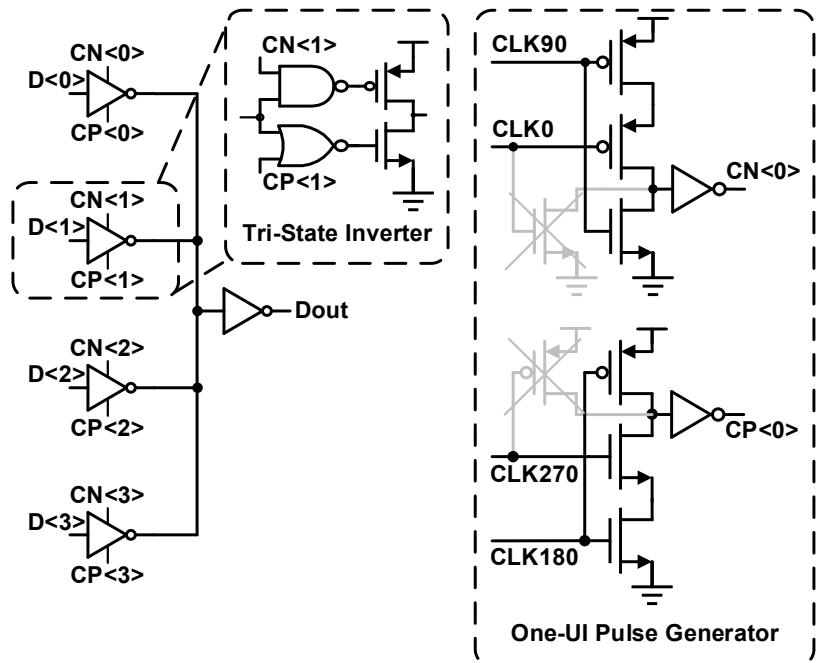

Fig. 10. Tri-state inverter-based 4:1 MUX and one-UI pulse generator.

The high-swing driver utilizes a stacked inverter configuration, with multiple transistors vertically aligned and middle devices, enabling a $2\times V_{DD}$ output swing, as depicted in Fig. 11 (a). Employing pseudo-differential topology can extend the swing to $4\times V_{DD}$. The output swing adjustment uses a binary encoding circuit, which results in an adjustable swing range of 1.32 to 3.2 $V_{PP}$ with 206 mV step value by independently controlling SW_P<0:3> and SW_N<0:3>, as shown in Fig. 11 (b). Edge-triggered pulses are incorporated at the gates of cascade transistors to mitigate overstress [27].

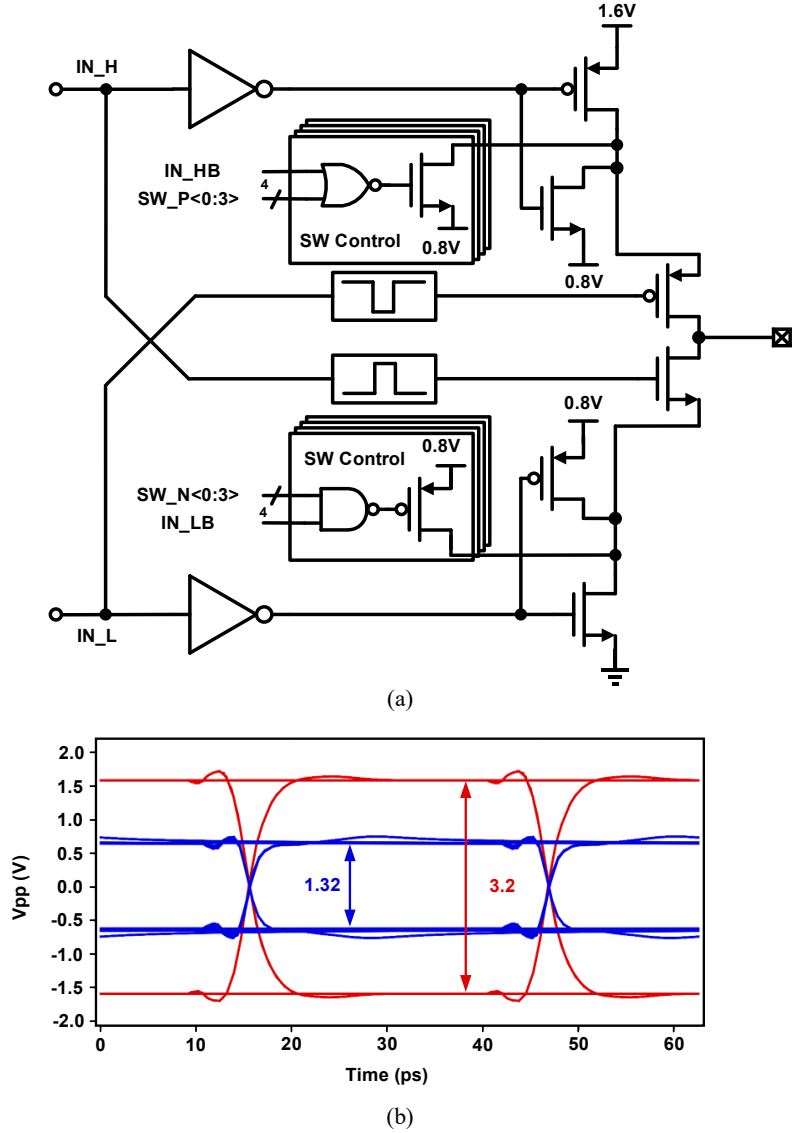

Fig. 11. (a) Schematic of the stacked inverter-based MRM driver with tunability. (b) Simulated eye diagram of the output differential signal.

## IV. Conclusion

In summary, we have proposed a novel three-segment silicon MRM and a matched drive circuit to realize an adjustable drive voltage of 1.32 V to 3.2 V, with a wider operating wavelength range to realize linear PAM-4 output, which ensures the IL of MRM has a greater range of adjustment, compared with traditional two-segment MRM. This work shows the potential of this novel high-linearity PAM-4 silicon micro-ring transmitter architecture with electronic-photonic hybrid DAC to be used in data centers, with a wider range of adjustable IL and ultra-high-linear PAM-4 output, which can be used for some key applications such as reducing link power consumption and bit error ratio. Besides, our proposed architecture is highly scalable and has some poti, for example, when the operating wavelength is shifted due to process deviation, the drive voltage can still be adjusted to achieve the linear output of PAM-4, or the optical peaking effect can also be used to improve the electro-optical bandwidth while ensuring the linear output of PAM-4 [28].